\documentclass[aps,pre,twocolumn,showpacs]{revtex4}
\usepackage{amsmath,amssymb,graphicx,bm}
\usepackage[none]{hyphenat}
\begin{document}

\title{Calculating the free energy difference by applying the Jarzynski equality to a virtual integrable system}
\author{Liyun Zhu}
\author{Jiao Wang}
\email{phywangj@xmu.edu.cn}
\affiliation{Department of Physics, Key Laboratory of Low Dimensional Condensed Matter Physics (Department of Education of Fujian Province), and Jiujiang Research Institute, Xiamen University, Xiamen 361005, Fujian, China}
\date{\today}

\begin{abstract}
The Jarzynski equality (JE) provides a nonequilibrium method to measure and calculate the free energy difference (FED). Note that if two systems share the same Hamiltonian at two equilibrium states, respectively, they share the same FED between these two equilibrium states as well. Therefore the calculation of the FED of a system may be facilitated by considering instead another virtual system designed to this end. Taking advantage of this flexibility and the JE, we show that by introducing an integrable virtual system, the evolution problem involved in the JE can be solved. As a consequence, FED is expressed in the form of an equilibrium equality, in contrast with the nonequilibrium JE it is based on. Numerically, this result allows FED to be computed by sampling the canonical ensemble directly and the computational cost can be significantly reduced. The effectiveness and efficiency of this scheme are illustrated with numerical studies of several representative model systems.
\end{abstract}


\maketitle

\sloppy{}

\section{Introduction}

The (Helmholz) free energy is a state variable of a thermodynamic system. When the system changes its state from one to another at the same temperature, the decrease of the free energy gives the largest work the system can output. As the free energy explains the phase behavior of a system and can be directly related to the experimentally determined properties, it plays an important role in a broad spectrum of applications~\cite{Chipot}.

Nevertheless, in general, to efficiently measure and calculate the free energy is challenging. According to the second law, the largest work can be captured only when the system changes its state reversibly, i.e., infinitely slow so that the process remains quasistatic. This makes the measurement of the free energy (the largest work) difficult, as any measurement has to be carried out in a reasonable, finite time. The numerical computation of the free energy is also difficult, because unlike ``mechanical'' state variables, which can be computed directly by sampling the equilibrium ensemble, the free energy involves the evaluation of the whole phase space by definition~\cite{Frenkel85, Frenkel}. A conventional method for computing the free energy difference (FED) between two given states is the thermodynamic integration method~\cite{Kirkwood}, by which one has to first compute some related state variables (e.g., the pressure, in an isothermal process) as a function of the medium equilibrium states of the quasistatic process that connects the two given states, then obtain FED by integrating this function. Obviously, this is computationally more expensive and inefficient than the computation of a mechanical state variable.

In 1997, Jarzynski found a significant equality that relates FED between two equilibrium states (at the same temperature) to the work done to the system in a nonequilibrium process~\cite{Jar97PRL, Jar97PRE}. Precisely, suppose the Hamiltonian of the system is $H({\mathbf s}; \lambda)$, where $\mathbf s$ is the system state and $\lambda$ is a system parameter. When the parameter is changed in time following a given prescribed protocol $\lambda(t)$ from $\lambda_{\rm A}$ at time $t_{\rm A}$ to $\lambda_{\rm B}$ at time $t_{\rm B}$, the Jarzynski equality (JE) states that
\begin{equation}
e^{-\beta\Delta F}=\langle e^{-\beta w}\rangle_{\mathrm A}.
\end{equation}
Here $\beta\equiv 1/(k_B T)$ is the inverse temperature, $\Delta F\equiv F_{\rm B}-F_{\rm A}$ is FED between equilibrium state A and B parameterized by $\lambda_{\rm A}$ and $\lambda_{\rm B}$, respectively, and $w$ is the work done to the system when it is evolved from an initial state sampled from the canonical ensemble of state A at time $t_{\rm A}$ up to time $t_{\rm B}$. The work depends on the initial condition; by repeating sampling of the initial condition, the work distribution can be established, over which the exponential work average can be evaluated and in turn FED is obtained. The angular brackets and the subscript A at the right-hand side (r.h.s.) of Eq.~(1) represent the average over the canonical ensemble of A. Note that the system does not necessarily relax to equilibrium state B at time $t_{\rm B}$, which is a profound property of the JE. Also note that when the system evolves, it can be isolated or coupled to the environment of temperature $T$~\cite{Jar97PRL, Jar97PRE, Jar04}.

Jarzynski's equality provides an alternative method for measuring and computing FED. As the time interval $t_{\rm B}-t_{\rm A}$ during which the system is driven can be finite and short, it seems particularly favorable for experimental measurements~\cite{exp1, exp2, exp3, exp4}. However, as pointed out by Jarzynski~\cite{Jar97PRL, Jar08} and other authors, in practice, to apply the JE directly may be inconvenient, because small work with rare probability weighs heavily for the exponential average $\langle e^{-\beta w}\rangle_{\mathrm A}$, a hefty sample could be needed to evaluate it accurately, and thus the cost could be demanding. Therefore, a key consideration in applying the JE directly is how to allocate the cost for sampling and driving the system. In general, for a given accuracy, the shorter the time interval $t_{\rm B}-t_{\rm A}$, the larger the work fluctuation and the sampling size needed. An empirical rule is to keep the work fluctuation less than $k_B T$~\cite{Dellago14}.

Since the JE was revealed, many efforts have been made to develop improved algorithms for computing FED. A thorough survey can be found in Ref.~\cite{Dellago14}. Roughly speaking, these efforts can be classified into two categories: one is to shorten the time needed to evolve the system by molecular dynamics simulations and another is to reduce the statistical uncertainty for evaluating
$\langle e^{-\beta w}\rangle_{\mathrm A}$. In the former, the main progress is the targeted free energy perturbation method developed by Jarzynski based on a generalized JE~\cite{Jar02}. This method is a variant of the free energy perturbation theory~\cite{Zwanzig}, which allows FED to be computed with crude trajectories simulated with large time steps~\cite{Dellago06, Dellago07}. To reduce the statistical uncertainty, the most ``straightforward'' way is to take the work biased sampling schemes to generate more trajectories whose work values dominate in calculating $\langle e^{-\beta w}\rangle_{\mathrm A}$. To this end, one way is to introduce an explicit bias function in calculating $\langle e^{-\beta w}\rangle_{\mathrm A}$ to enhance the sampling of important trajectories~\cite{Ytreberg, Athenes} and another is to introduce a parameter that biases the contribution of different trajectories to make sure that all their contributions are fully taken into account~\cite{Sun03, Sun04}. The latter can be viewed as a thermodynamic integration procedure in trajectory space~\cite{Dellago14}. For enhancing sampling of important trajectories, general methods designed for simulating rare events, e.g., the population dynamics with cloning~\cite{Giardina}, might be adopted as well. In order to reduce the statistical uncertainty, another important direction to explore is to optimize the protocol. Note that the JE does not depend on the details of the protocol; all paths from $\lambda_{\rm A}$ to $\lambda_{\rm B}$ give the same result of FED. But the work distribution depends on the protocol, implying the existence of an optimal protocol that can minimize the work fluctuation. If the changing rate of $\lambda$ is small, example studies suggest that a protocol with small mean work also leads to small statistical uncertainty~\cite{Jar06, Sun04}. Considering this, Schmiedl and Seifert found that an optimal protocol may consist of two jumps at $t_{\mathrm A}$ and $t_{\mathrm B}$~\cite{Tim07}.

In fact, the flexibility implied by the JE lies not only in the protocol; the dynamics of the system can be manipulated as well. For example, the JE can be generalized to incorporate an artificial flow field to escort a trajectory such that in the best situations, it may give FED exactly by sampling the initial condition and evolving the system only once~\cite{Jar08}. The drawback of this scheme, however, is that it is hard to solve the appropriate flow field except in some special cases~\cite{Jar08}.

Recently, Gong's group studied the general methods to suppress the work fluctuation for a given protocol by applying a control field to the system~\cite{JB13, JB14}. The applied control field is expressed as an additional term to the Hamiltonian, which is turned off before time $t_{\rm A}$ and after time $t_{\rm B}$ but turned on for $t_{\rm A}<t<t_{\rm B}$. For an integrable system, based on the shortcuts to adiabatic process, the authors worked out the control field that makes the work distribution identical to that of quasistatic processes from A to B~\cite{JB13}. Hence the work fluctuation is suppressed to be the minimum allowed in principle. Later this scheme was generalized to non-integrable systems where the control field is determined by the optimal control technique~\cite{JB14}. In this general scheme, minimizing the fluctuation of $e^{-\beta w}$ from its average $e^{-\beta \Delta F}$ [see Eq.~(1)] has been taken as the explicit control target, hence it can be adopted as a boosting JE method for evaluating FED for both experimental and numerical studies.

In this work we explore a different strategy for boosting the calculation of FED based on the JE. We also take advantage of the fact that the dynamics of the system can be manipulated, but unlike in Refs.~\cite{JB13, JB14}, we get rid of the original Hamiltonian of the system during the time interval $t_{\rm A}<t<t_{\rm B}$ but replace it with an integrable dynamics such that the evolution of the system can be solved analytically. As a result, an $equilibrium$ equality of FED, in contrast with the underlying nonequilibrium JE, is derived. Numerically, this equilibrium equality allows FED to be computed like a mechanical state variable~\cite{Frenkel85, Frenkel} by sampling the canonical ensemble directly, which is a significant simplification. Compared with the direct JE algorithm, the computational cost can be saved for orders in the studied examples. In the following, we will first outline the general scheme of our strategy, then apply it to the protocol that the system changes its volume from state A to B. The analytical results will be checked with numerical examples and extended to more general protocols. Finally, some related issues will be discussed with a brief summary.

\section{A general scheme: Applying the JE to a virtual integrable system}

Our task is to calculate the FED of the system $H({\mathbf s}; \lambda)$ between states A and B. Consider a different Hamiltonian system $\tilde H({\mathbf s}; \Lambda)$ that shares the same phase space, where $\Lambda$ represents its parameter set. If, for a certain value of $\Lambda$, denoted as $\Lambda_{\rm A}$, this Hamiltonian is identical to $H({\mathbf s}; \lambda_{\rm A})$, i.e., $\tilde H({\mathbf s}; \Lambda_{\rm A})=H({\mathbf s}; \lambda_{\rm A})$, then the two systems share the same equilibrium distribution $P_{\rm A}({\mathbf s})\equiv e^{-\beta \tilde H({\mathbf s}; \Lambda_{\rm A})}/Z_{\rm A}=e^{-\beta H({\mathbf s}; {\lambda}_{\rm A})}/Z_{\rm A}$ and therefore the same free energy $\tilde F_{\rm A}=F_{\rm A}=-\ln Z_{\rm A}/\beta$. Here $Z_{\rm A}$ is the partition function of their common state A. Similarly, if for $\Lambda_{\rm B}$ we have $\tilde H({\mathbf s}; \Lambda_{\rm B})=H({\mathbf s}; \lambda_{\rm B})$, then the two systems have the same free energy $\tilde F_{\rm B}=F_{\rm B}=-\ln Z_{\rm B}/\beta$ at state B as well. Given these, the FED of the original system $\Delta F=F_{\rm B}-F_{\rm A}$ is equal to that of $\tilde H$, $\Delta \tilde F=\tilde F_{\rm B}-\tilde F_{\rm A}$, and therefore can be calculated by the JE with $\tilde H$ instead:
\begin{equation}
e^{-\beta \Delta F}=e^{-\beta \Delta \tilde F}=\langle e^{-\beta \tilde w}\rangle_{\mathrm A}.
\end{equation}
Here $\tilde w$ is the work performed on the ``virtual'' system $\tilde H$ when it is driven by the control parameter set $\Lambda$ from $\Lambda_{\rm A}$ to $\Lambda_{\rm B}$ with a given protocol $\Lambda(t)$. This relation has been pointed out and utilized in Refs.~\cite{JB13, JB14}, which is very flexible: It gives us the freedom to manipulate not only the protocol, but also the Hamiltonian. We emphasize that the only requirements are
\begin{equation}
\tilde H({\mathbf s}; \Lambda_{\alpha})= H({\mathbf s}; {\lambda}_{\alpha}), ~~\alpha={\rm A}, {\rm B}.
\end{equation}
At other system parameter values, the two Hamiltonians can be different and arbitrary.

\begin{figure}[!]
\vskip-0.2cm
\includegraphics[width=8.8cm]{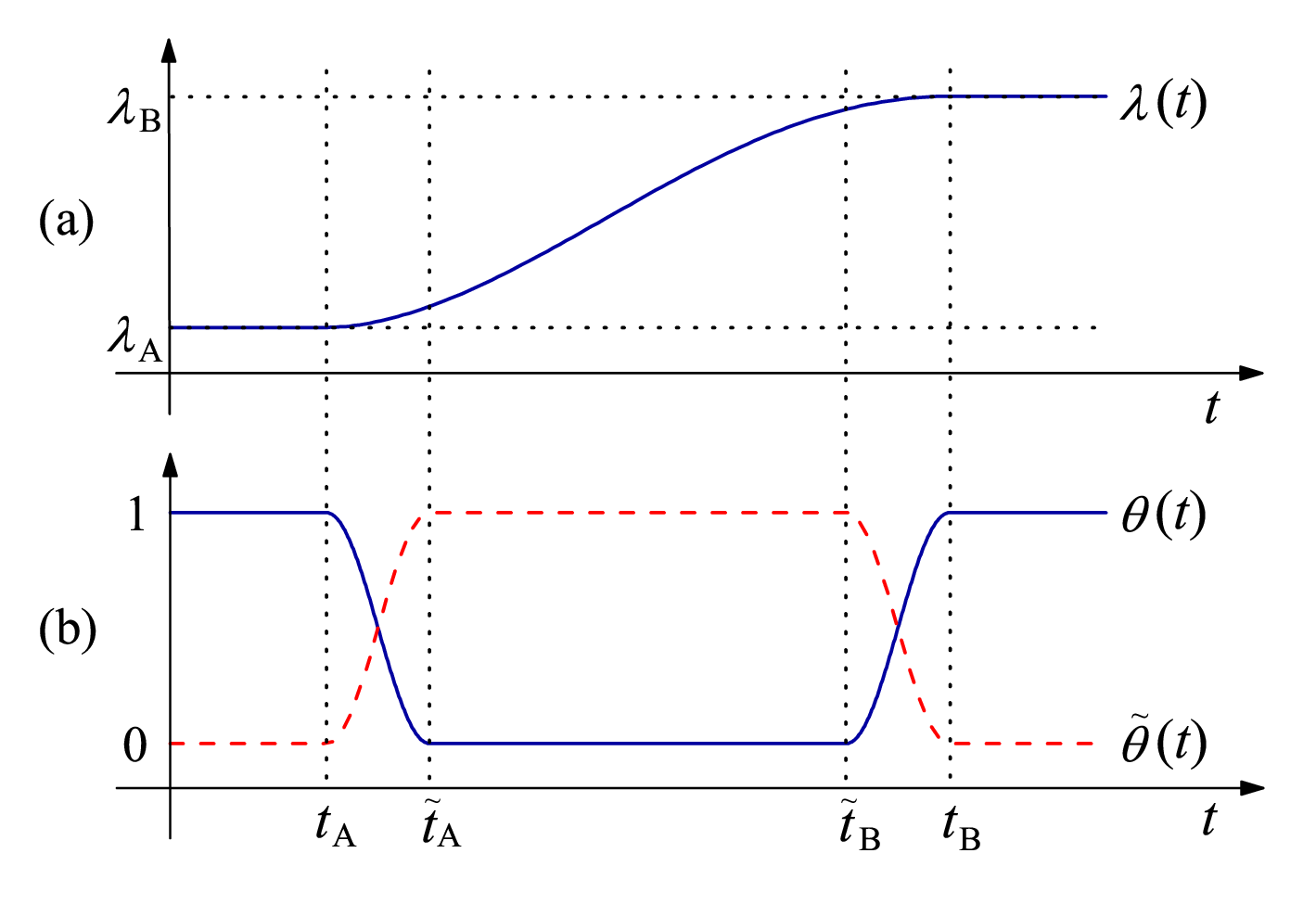}
\vskip-0.2cm
\caption{Schematic plot of the protocol adopted in the scheme based on the JE (a) and in our suggested scheme (b) for evaluating the free energy difference. The two switch functions $\theta$ and $\tilde \theta$ introduced in our scheme are used to suppress the original interaction but activate a virtual interaction for $\tilde t_{\rm A}<t<\tilde t_{\rm B}$ (and vice versa for $t\le t_{\rm A}$ and $t \ge t_{\rm B}$). Protocol $\tilde \lambda (t)$ in our scheme (not shown) is arbitrary given that $\tilde \lambda(t)=\lambda_{\rm A}$ for $t\le t_{\rm A}$ and $\tilde \lambda(t)=\lambda_{\rm B}$ for $t\ge t_{\rm B}$. }
\end{figure}

In the following we will show that, indeed, this scenario can lead to significant simplification in calculating $\Delta F$. Suppose that the system consists of $N$ particles and its Hamiltonian is
\begin{equation}
H({\mathbf s}; \lambda)=\sum \frac{\mathbf{p}_i^2}{2m_i} + U(\mathbf{r};\lambda),
\end{equation}
where $m_i$, $\mathbf{r}_i$, and $\mathbf{p}_i$ are, respectively, the mass, position, and momentum of the $i$th particle, and $\mathbf{s}=(\mathbf{p},\mathbf{r})$ with $\mathbf{p}\equiv (\mathbf{p}_1,\cdots, \mathbf{p}_N)$ and $\mathbf{r}\equiv (\mathbf{r}_1,\cdots, \mathbf{r}_N)$. To apply the JE, the protocol should follow that $\lambda(t)=\lambda_{\rm A}$ for $t\le t_{\rm A}$ and $\lambda(t)=\lambda_{\rm B}$ for $t\ge t_{\rm B}$ [see Fig.~1(a)]. When the protocol is assigned, $\Delta F$ can be obtained by the JE directly.

Alternatively, we can obtain $\Delta F$ in the following virtual system by using Eq.~(2):
\begin{equation}
\tilde H ({\mathbf s}; \Lambda)=\sum \frac{\mathbf{p}_i^2}{2m_i} + \theta U(\mathbf{r};\tilde \lambda)+ \tilde \theta V(\mathbf{r};\tilde \lambda).
\end{equation}
Here $\Lambda=(\theta, \tilde \theta, \tilde \lambda)$, where $\theta$ and $\tilde \theta$ are two switch functions. In order to ensure that at $t_{\rm A}$ and $t_{\rm B}$ the two Hamiltonians are identical, we set $\theta$, $\tilde \theta$, and $\tilde \lambda$ as follows: For $t \le t_{\rm A}<\tilde t_{\rm A}$ and $t \ge t_{\rm B}>\tilde t_{\rm B}$, we assign $\theta=1$ and $\tilde \theta=0$ to adopt the interaction, $U$, of the original system. In addition, we assume that $\tilde \lambda (t)=\lambda(t)$ for $t\le t_{\rm A}$ and $t\ge t_{\rm B}$. With these settings, Eq.~(3) is guaranteed to hold; $\Delta F$ of the original system is therefore identical to that of the virtual system and can thus be obtained with the latter.

But for $\tilde t_{\rm A}<t<\tilde t_{\rm B}$, we set $\theta=0$ and $\tilde \theta=1$ instead, to switch the interaction to the introduced virtual interaction, $V$ [see Fig.~1(b)]. It is worth noting that, in principle, any $V$ allowed by physics is acceptable. Moreover, the protocol $\tilde \lambda(t)$ can be arbitrary over $t_{\rm A} < t < t_{\rm B}$, as long as it changes from $\lambda_{\rm A}$ at $t=t_{\rm A}$ to $\lambda_{\rm B}$ at $t=t_{\rm B}$. These flexibilities and freedoms are the advantages the introduced virtual system brings, and our main motivation in this work is to make use of them to facilitate the calculation of FED.

Before proceeding, we notice that by taking the limits $\tilde t_{\rm A}\to t_{\rm A}$ and $\tilde t_{\rm B}\to t_{\rm B}$, we can write down part of the work immediately. As the Hamiltonian changes abruptly at $t_{\rm A}$ and $t_{\rm B}$, the work done to the system is~\cite{Jar97PRL}, respectively,
\begin{align}
\tilde w_{\rm A}\equiv \Delta \tilde H|_{\tilde t_{\rm A}\to t_{\rm A}} &=V({\mathbf r}(t_{\rm A}); \lambda_{\rm A})-U({\mathbf r}(t_{\rm A}); \lambda_{\rm A}); \nonumber \\
\tilde w_{\rm B}\equiv \Delta \tilde H|_{\tilde t_{\rm B}\to t_{\rm B}} &=U({\mathbf r}(t_{\rm B}); \lambda_{\rm B})-V({\mathbf r}(t_{\rm B}); \lambda_{\rm B}).
\end{align}
Following Eq.~(2), we then have
\begin{equation}
e^{-\beta \Delta F}=\langle e^{-\beta (\tilde w_{\rm A}+\tilde w_{\rm B}+\tilde w_{V})}\rangle_{\mathrm A},
\end{equation}
where $\tilde w_{V}$ is the work done to the virtual system with the introduced interaction $V({\mathbf r}; \tilde \lambda)$ when being driven by $\tilde \lambda$ from $\tilde \lambda= \lambda_{\rm A}$ to $\tilde \lambda=\lambda_{\rm B}$.

One advantage of this scheme is apparent now: In principle, for an integrable interaction $V$, $\tilde w_{V}$ can be solved; then the calculation of FED reduces to an equilibrium average without any explicit nonequilibrium quantities. Numerically, as evolving the system is avoided, the reduction of the simulation cost is guaranteed.

\section{Free energy difference between two volumes}

As an application of our general scheme, here we discuss the FED of a system at two different volumes. The derivation of FED between two values of any other parameter or parameter set is similar (see Sec.~V). For the sake of simplicity, we consider one-dimensional (1D) systems in this section. The possible extension to two-dimensional (2D) and three-dimensional (3D) cases will be discussed in Sec.~VI.

For a 1D system, $\mathbf{r}=\mathbf{x}\equiv (x_1,\cdots, x_N)$ and $\mathbf{p}=(p_1,\cdots, p_N)$, where $x_i$ and $p_i$ are the position and the momentum of the $i$th particle. Its volume is the length of the system, denoted as $L$. By the JE, we can take the protocol, identifying $\lambda$ with $L$, as follows: At $t_{\rm A}$, the system volume is $L_{\rm A}$; then we press or pull one end of the system at a fixed velocity $u$ to make its volume $L_{\rm B}$ at $t_{\rm B}=t_{\rm A}+(L_{\rm B}-L_{\rm A})/u$. During this process the system keeps its interaction $U({\mathbf x}, L(t))$. By our scheme with the virtual system, the key difference is that at $t_{\rm A}$, we replace $U$ by the virtual potential $V$, and at $t_{\rm B}$, we switch back to $U$. For our aim here one convenient option of $V$ is that which consists of $N_c$ identical cells of hard walls (see Fig.~2). We set $N_c$ large enough to make sure that in each cell there is at most one particle, so that the particles become noninteractive. At $t=t_{\rm A}$, we press or pull one boundary of each cell with velocity $u$ as well until time $t_B$, during which when a particle collides with any boundary of its cell, it is reflected back elastically. The work $\tilde w_{V}$ done to the system can thus be obtained by summing up the work done to each particle by the moving boundary of its cell, denoted as $\tilde w_{V,i}$, which can be solved analytically (Eq.~(A8) in Appendix A; see also Ref.~\cite{Lua05}). The advantage of the adopted $V$ is that it keeps the order of particles. This is particularly crucial for a lattice, otherwise the original interaction $U$ may not be retrieved at time $t_{\rm B}$.

\begin{figure}[!]
\vskip-0.2cm
\includegraphics[width=8.8cm]{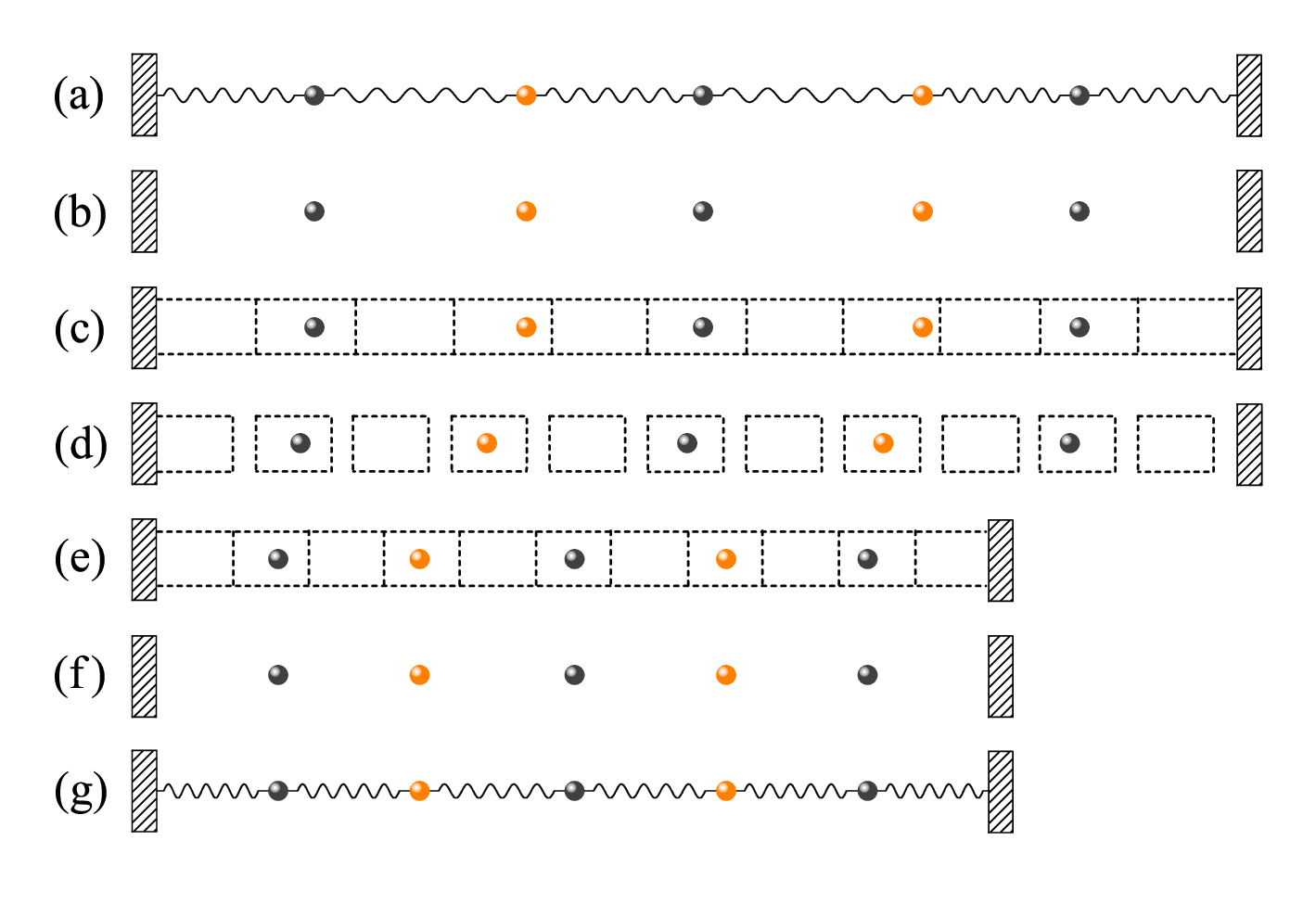}
\vskip-0.2cm
\caption{Illustration of the suggested scheme for evaluating the free energy difference when the system has a reference system volume, $L_{\rm A}$ (a), and a given system volume, $L_{\rm B}$ (g), with a 1D diatomic lattice as illustrating example. (a) For $t\le t_{\rm A}$, the original interaction $U$, represented by wavy lines, operates. (b) At $t=t_{\rm A}$, interaction $U({\mathbf x}(t_{\rm A}); L_{\rm A})$ is cut off and (c) the virtual auxiliary interaction, $V({\mathbf x}(t_{\rm A}); L_{\rm A})$, represented by cells, is switched on simultaneously. At this time work $\tilde w_{\rm A}$ is calculated. (d) For $t_{\rm A}<t<t_{\rm B}$, each particle is ``pressed'' by the right boundary of its cell moving at velocity $u$. Meanwhile work $\tilde w_{V}$ is evaluated. (e) At $t=t_{\rm B}$, cells are aligned one by one, then (f) interaction $V({\mathbf x}(t_{\rm B}); L_{\rm B})$ is removed and (g) the original interaction $U({\mathbf x}(t_{\rm B}); L_{\rm B})$ is activated again. At this moment work $\tilde w_{\rm B}$ is evaluated.}
\end{figure}

It is rewarding to take the limits $N_c \to \infty$ and $u \to 0$ further, following which we have immediately $x_i(t_{\rm B})=r x_i(t_{\rm A})$ with $r\equiv L_{\rm B}/L_{\rm A}$ and $\tilde w_{V,i}=(1/r^2-1) p_i^2 /(2m_i)$ (Eq.~(A9) in Appendix A), allowing Eq.~(7) to be rewritten as
\begin{equation}
e^{-\beta \Delta F}=r^N\langle e^{\beta [U({\mathbf x}; L_{\rm A})-U(r{\mathbf x}; L_{\rm B})]}\rangle_{{\mathrm A},{\mathbf x}}
\end{equation}
with the distribution function for averaging $P_{{\rm A},{\mathbf x}}\equiv e^{-\beta U({\mathbf x}; L_{\rm A})}/Z_{{\rm A},{\mathbf x}}$ and $Z_{{\rm A},{\mathbf x}}=\int  e^{-\beta U({\mathbf x}; L_{\rm A})}d{\mathbf x}$. Here the prefactor $r^N$ at the r.h.s. is for the result of $\langle e^{-\beta \tilde w_{V}}\rangle_{\mathrm A}$, which can be integrated out independently from $\langle e^{-\beta (\tilde w_{\rm A}+\tilde w_{\rm B})}\rangle_{\mathrm A}$ as $\tilde w_{V}$ depends only on variable $\mathbf p$ while $\tilde w_{\rm A}$ and $\tilde w_{\rm B}$ depend only on $\mathbf x$. The exponential average at the r.h.s. of Eq.~(8) corresponds to $\langle e^{-\beta (\tilde w_{\rm A}+\tilde w_{\rm B})}\rangle_{\mathrm A}$. The derivation of Eq.~(8) and its extension to 2D and 3D cases is detailed in Appendix B.

Theoretically, this result reveals a new equilibrium relation between the free energy of a system at two different volumes. It is interesting in view of the fact that it is derived from the JE that is established based on nonequilibrium processes. Numerically, the standard Monte Carlo algorithm involving variable $\mathbf x$ only can be applied directly. In doing so, as the exponential average of $\tilde w_{\rm A}+\tilde w_{\rm B}$, rather than that of $\tilde w=\tilde w_{\rm A}+\tilde w_{\rm B}+\tilde w_V$, is evaluated, for a given accuracy the ensemble size can be reduced because the distribution of $\tilde w_{\rm A}+\tilde w_{\rm B}$ is narrower than that of $\tilde w_{\rm A}+\tilde w_{\rm B}+\tilde w_V$. This simplifies the computation of FED further.

\begin{figure}[!t]
\vskip-0.2cm
\includegraphics[width=8.8cm]{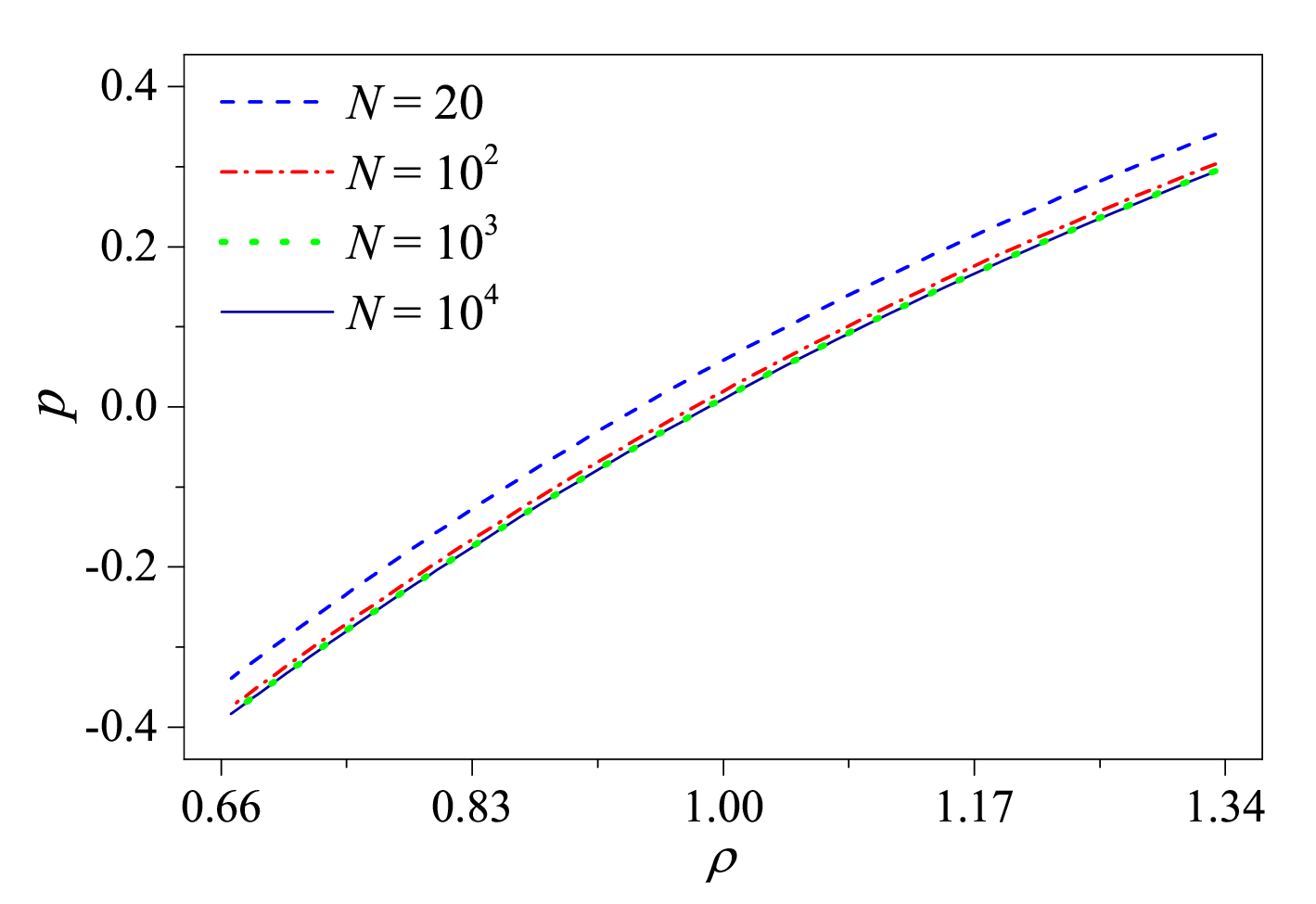}
\vskip-0.2cm
\caption{The pressure of the diatomic Toda lattice of $N$ particles as a function of the particle density. $\beta=50$ here and in Figs.~4 and 5. }
\end{figure}

\section{Free energy difference of two one-dimensional models}

To test the effectiveness and efficiency of our main results Eqs.~(7) and (8), here we study two representative model systems as examples. Note that in all the figures (Figs.~3-7) where our numerical results are provided, the statistical uncertainty of the data (``error bar'') is smaller than at least one-tenth of the thickness of the line, or the size of the symbols that represent them, and hence is not shown.

The first model is the one-dimensional (1D) diatomic Toda lattice~\cite{Hatano} with
\begin{equation}
U=\sum [e^{-(x_{i+1}-x_i-1)}+(x_{i+1}-x_i-1)].
\end{equation}
The two kinds of particles have mass 1 and 2 and align alternately. Note that this model is non-integrable~\cite{Shunda}. The fixed boundary conditions are taken by fixing the zeroth and the $(N+1)$th particle at the left and right boundary. For our aim here we also calculate the FED with the conventional thermodynamic integration method~\cite{Kirkwood} and use the result as a benchmark. To this end, the pressure of the system as a function of the system size, or equivalently, the particle density $\rho\equiv N/L$, is calculated with high accuracy by using the canonical ensemble Monte Carlo algorithm (see Fig.~3). The free energy difference is then obtained by integrating the pressure based on the relation $({\partial F}/{\partial V})_{N,T}=-p$. The statistical uncertainty of the simulated pressure is smaller than $2\times 10^{-6}$ and the corresponding uncertainty of FED per particle, $\Delta f\equiv\Delta F/N$, is less than $10^{-5}$ (see the dashed and the solid line in Fig.~4).

\begin{figure}[!t]
\vskip-0.2cm
\includegraphics[width=8.8cm]{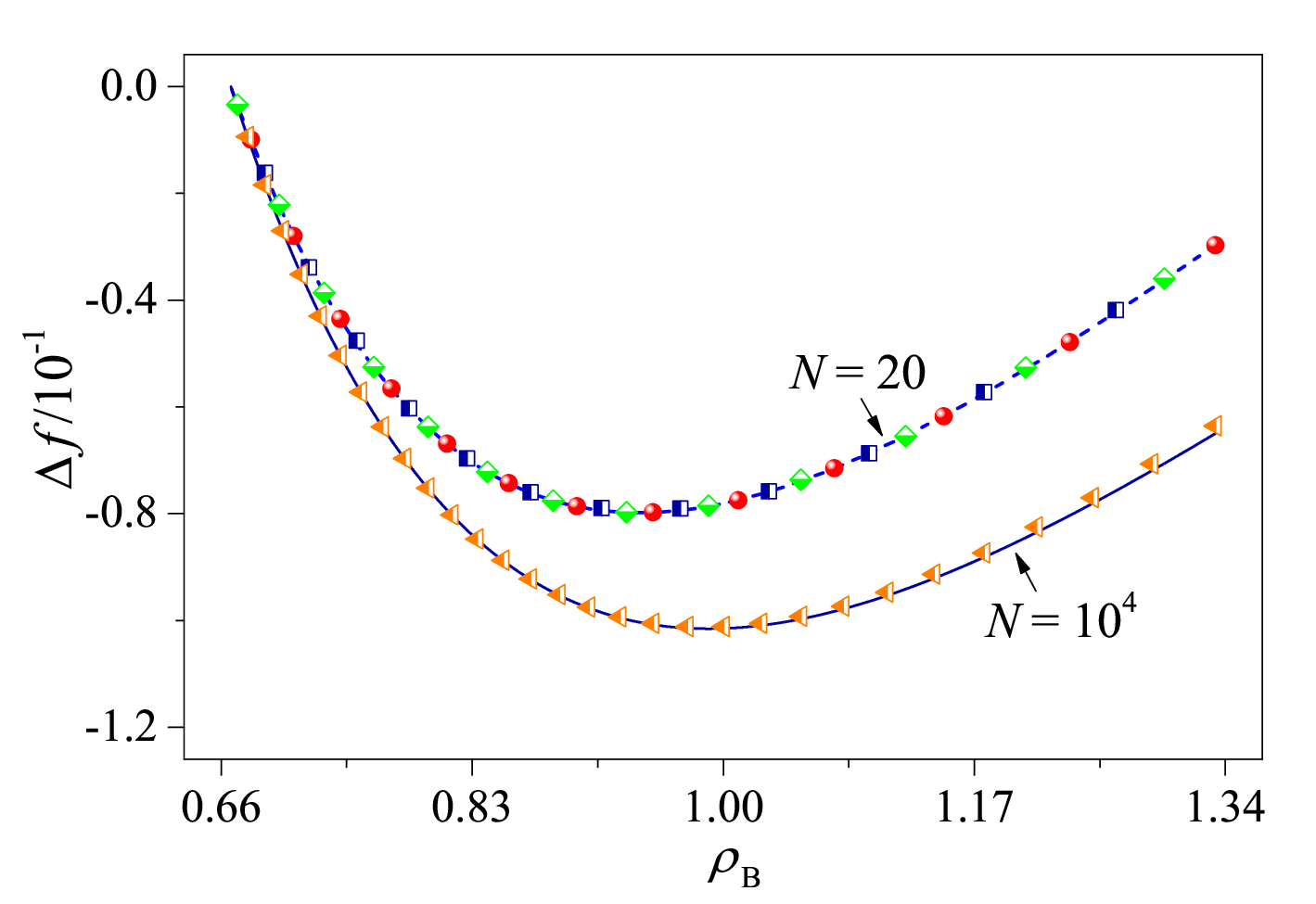}
\vskip-0.2cm
\caption{The free energy difference per particle of the 1D diatomic Toda lattice between system volume $L_{\rm A} = 3N/2$ and a given volume $L_{\rm B}=N/\rho_{\rm B}$ that changes from $L_{\rm A}$ to $L_{\rm A}/2$. The squares and the diamonds are for the direct JE method and our scheme Eq.~(7) with $N_c=300$, respectively, for $N=20$ with $u=0.1$ and the average ensemble size $10^5$. The dots (triangles) are for our scheme Eq.~(8) for $N=20$ ($N=10^4$) with the average ensemble size $10$. The dashed (solid) line gives the result of the conventional method by integrating the pressure [see Fig.~(3)] for $N=20$ ($N=10^4$).}
\end{figure}

The results of FED computed by using the direct the JE method, and our method with Eqs.~(7) and (8), respectively, are compared in Fig.~4. For all three methods, the involved average ensemble of microscopic states of state A (with volume $L_{\rm A}$) are generated by the canonical ensemble Monte Carlo algorithm. For the direct JE method, the sampled microscopic states are set to be the initial states and evolved by the double precision, fourth order Runge-Kutta algorithm with the time step $h=10^{-3}$. For $N=20$ with $u=0.1$ and the average ensemble size $10^5$, the relative deviation from the benchmark of the results by the direct JE method is less than $0.9\%$. For the same settings, our method based on Eq.~(7) gives the same accurate results, but as $\tilde w_V$ has been solved analytically, the simulation time is only about $3\times 10^{-3}$ of the former.

The most efficient one is our method based on Eq.~(8). To reach the same accuracy, it needs only ten samples. So not only the time for evolving the system is completely saved, but also the cost for sampling is reduced remarkably. Indeed, as expected and shown in Fig.~5, the distribution of $\tilde w_{\rm A}+\tilde w_{\rm B}$ involved in Eq.~(8) is much narrower than that of $w$ involved in the direct JE method. As a comparison, for $N=20$ the computation time this scheme takes is only about $3\times 10^{-7}$ of that by the direct JE method. It is so efficient that it can be applied to a much bigger system (e.g., $N=10^4$; see Fig.~4) where the computational cost for the direct JE method has been forbiddingly expensive.

The second example is a gas model with repulsive interaction
\begin{equation}
U=\sum (x_{i+1}-x_i)^{-6}.
\end{equation}
All particles have a unity mass and the fixed boundary conditions are assumed as well. All the simulation details are the same as in the first example. In Fig.~6, the results of FED by our scheme with Eq.~(8) are compared with those by the direct JE method and by another method based on Eq.~(11) (see the following). Note that the systematically biased deviation of the latter two methods at larger particle density is due to insufficient sampling, which has been confirmed by changing the average ensemble size.

\begin{figure}[!t]
\vskip-0.2cm
\includegraphics[width=8.8cm]{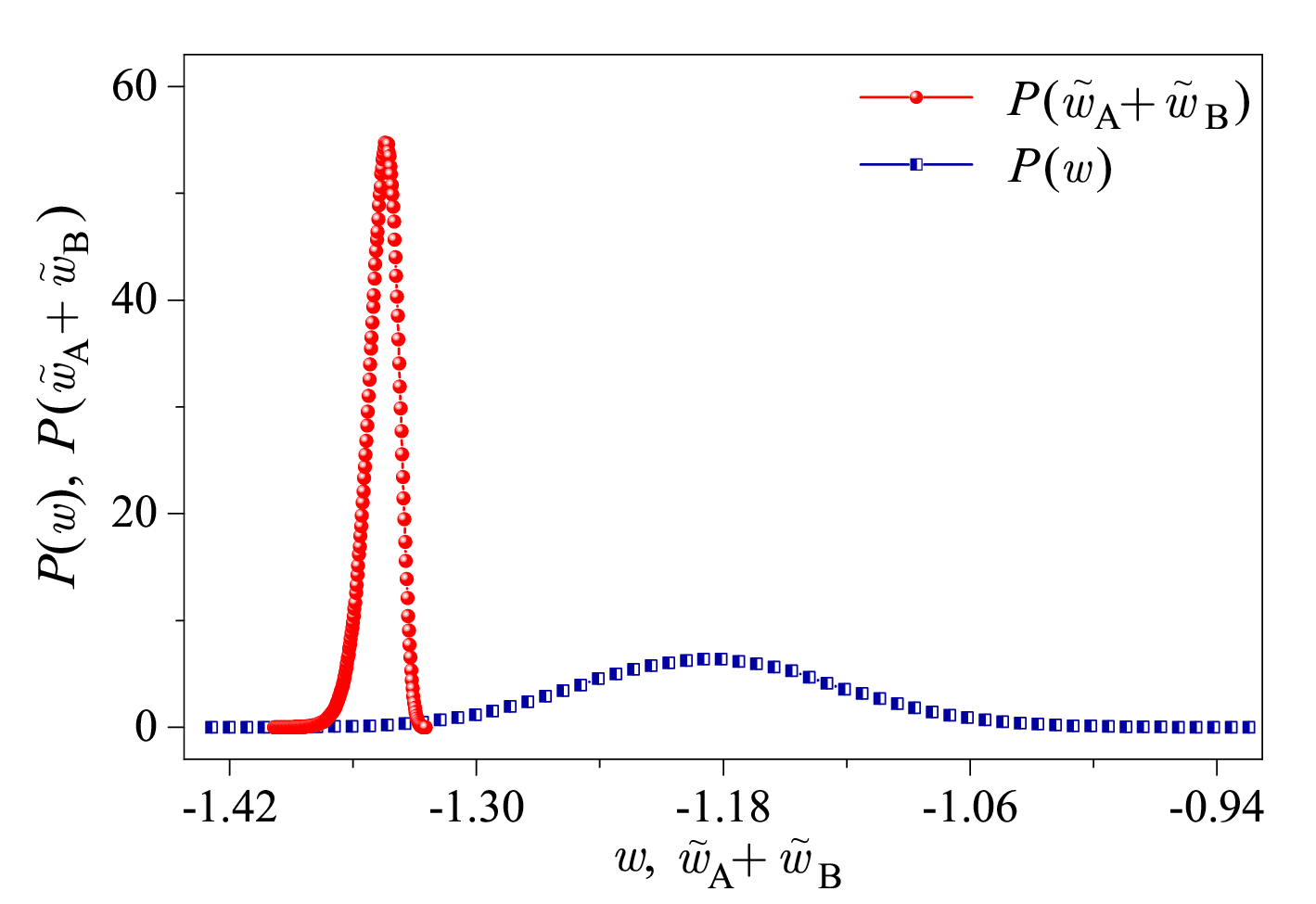}
\vskip-0.2cm
\caption{Comparison of the work distribution involved in our scheme based on Eq.~(8) (dots) and that in the direct JE method with $u=0.1$ (squares) for the diatomic Toda lattice of $N=20$. The initial and final system length is $L_{\rm A}=30$ and $L_{\rm B}=25$ [$\rho_{\rm B}=0.8$; see Fig.~(4)], respectively.}
\end{figure}

For the gas of identical particles where their position order is irrelevant, the FED between two system volumes can be calculated in a different way. Let us consider the following four systems, all consisting of $N$ particles of the same mass: (1) the system size is $L_{\rm A}$ and the interaction is $U$; (2) the system size is $L_{\rm A}$ but without interaction; (3) the system size is $L_{\rm B}$  without interaction; and (4) the system size is $L_{\rm B}$ and the interaction is $U$. System (2) and (3) are actually ideal gases. Obviously, the FED we want is in fact that between system (1) and system (4), i.e., $\Delta F=\Delta F_{14}=F_4-F_1$ ($F_i$ is the free energy of the $i$th system), which can be expressed in a chain as $\Delta F=\Delta F_{12}+\Delta F_{23}+\Delta F_{34}$. On the one hand, $F_{12}$ and $F_{34}$ can be obtained by the free energy perturbation theory~\cite{Zwanzig} or equivalently as the limiting result of the JE (see Eq.~(5) in Ref.~\cite{Jar97PRL}), which read $\Delta F_{12}=-\ln \langle e^{\beta U}\rangle_{\rm A}/\beta$ and $\Delta F_{34}=\ln \langle e^{\beta U}\rangle_{\rm B}/\beta$, respectively. On the other hand, as the partition function of an ideal gas is known, the FED between the ideal gases (2) and (3) can be written down straightforwardly: $\Delta F_{23}=-N\ln (L_{\rm B}/L_{\rm A})/\beta=-N\ln r/\beta$. As a consequence, we have
\begin{equation}
e^{-\beta \Delta F}=r^N[\langle e^{\beta U({\mathbf x}; L_{\rm A})}\rangle_{{\mathrm A},{\mathbf x}}/\langle e^{\beta U({\mathbf x}; L_{\rm B})}\rangle_{{\mathrm B},{\mathbf x}}].
\end{equation}
Comparing with Eq.~(8), an essential difference is that another ensemble average with $P_{{\rm B}, {\mathbf x}} = e^{-\beta U({\mathbf x}; L_{\rm B})}/Z_{{\rm B},{\mathbf x}}$ and $Z_{{\rm B}, {\mathbf x}} = \int e^{-\beta U({\mathbf x}; L_{\rm B})} d{\mathbf x}$, is involved here. For the gas model under study, the algorithm based on Eq.~(11) is not as efficient as that based on Eq.~(8), either, although it is more efficient than the direct JE method where evolving the system is avoided.

\begin{figure}[!t]
\vskip-0.2cm
\includegraphics[width=8.8cm]{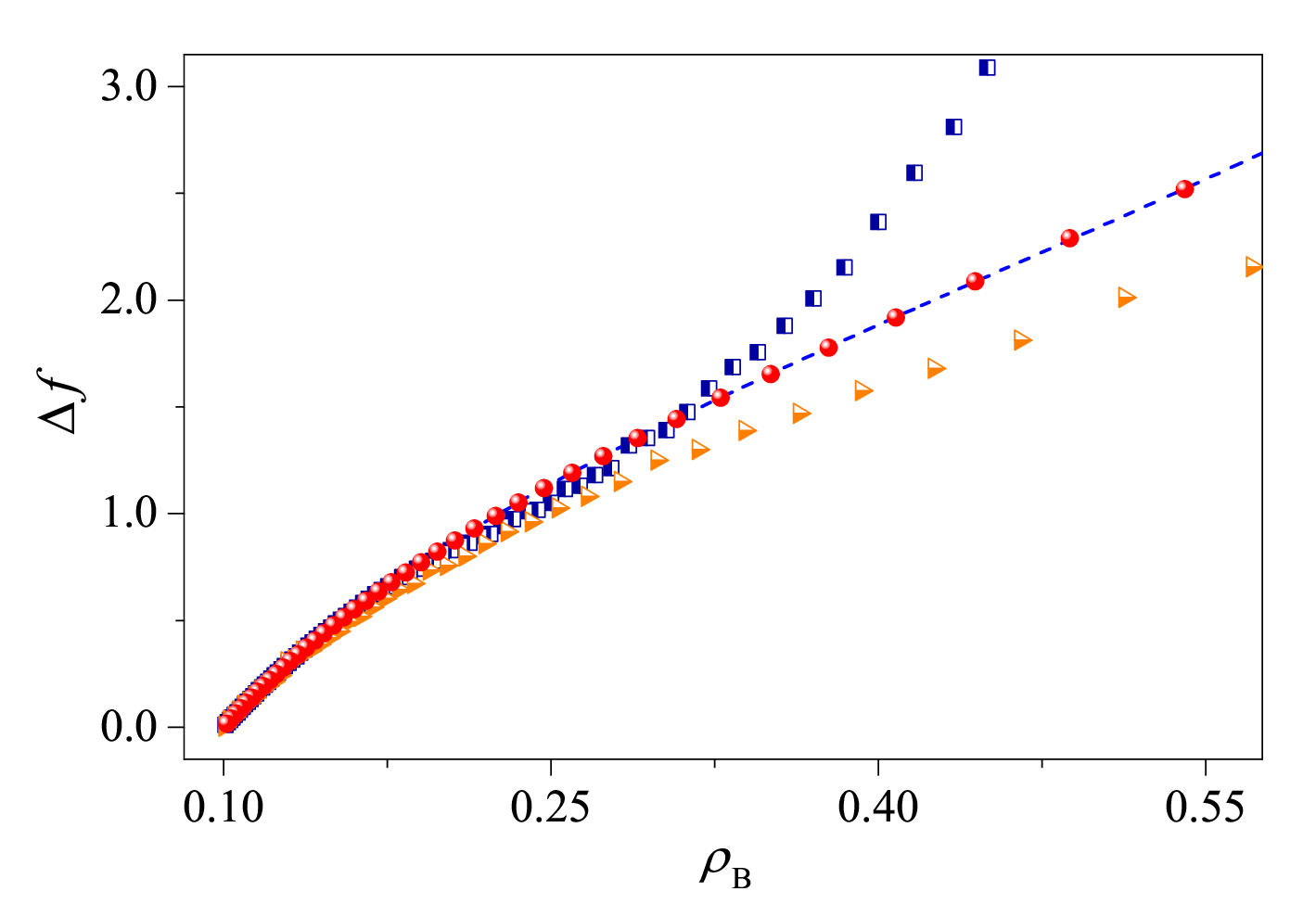}
\vskip-0.2cm
\caption{The free energy difference per particle of the gas model ($N=20$ and $\beta=1$) between system volume $L_{\rm A} = 10N$ and $L_{\rm B}=N/\rho_{\rm B}$. Squares, dots, and triangles are for, respectively, the results by the direct JE method ($u=0.1$), our scheme with Eq.~(8), and that based on Eq.~(11). For all three cases the average ensemble size is $10^4$. The dashed line is for the conventional method by integrating the numerically obtained pressure (not shown).}
\end{figure}

\section{Free energy difference between two general states}

As shown in Sec.~II, not only for the FED between two volumes, our general scheme based on Eq.~(7) is equally applicable to the FED between two states determined by other parameters as well. The key task is to design the virtual potential to facilitate the calculation of $\tilde w_V$. This can be fulfilled by cutting interactions to make particles move independently, just as we have done by introducing the hard-wall-cell potential. In principle, as the motion of each particle is a one-body problem, it is integrable and can be solved definitely. To this end, the hard-wall-cell potential is only one option. If the considered parameter is not the volume, another feasible choice could be an onsite harmonic potential array that confines each particle to move around its equilibrium position. For numerical calculations, for a given parameter a better choice of the virtual potential should be one that makes the distribution of $\tilde w_{\rm A} + \tilde w_{\rm B} + \tilde w_V$ narrower so that the sampling cost is less. To this end, an appropriate protocol can help additionally. For example, assuming $t_{\rm B}-t_{\rm A}\to \infty$ will not add any more computational cost as $\tilde w_V$ can be solved analytically, but it may suppress the fluctuations of $\tilde w_V$ and $\tilde w_{\rm A} + \tilde w_{\rm B} + \tilde w_V$.

If the system state is parameterized by a set of parameters $\Gamma$ to which the volume does not belong, the FED between two states A and B can be obtained by the free energy perturbation theory~\cite{Zwanzig}:
\begin{align}
e^{-\beta \Delta F}&=\langle e^{\beta [H({\mathbf s}; \Gamma_{\rm A})-H({\mathbf s}; \Gamma_{\rm B})]} \rangle_{\mathrm A} \nonumber\\
&=\langle e^{\beta [U({\mathbf x}; \Gamma_{\rm A})-U({\mathbf x}; \Gamma_{\rm B})]}\rangle_{{\mathrm A},{\mathbf x}}
\end{align}
This result can be derived from the JE with a limiting protocol that $\Gamma$ changes instantaneously from $\Gamma_{\rm A}$ to $\Gamma_{\rm B}$~\cite{Jar97PRL}. As ${\mathbf x}$ remains unchanged, it cannot be applied when the volume change is involved.

However, taking our scheme, Eq.~(12) can be extended straightforwardly to incorporate the volume change as follows: At time $t_{\rm A}$, the potential $U({\mathbf x}; \Gamma_{\rm A}, L_{\rm A})$ is switched off and the hard-wall-cell potential is switched on; Next, the volume is changed from $L_{\rm A}$ to $L_{\rm B}$ following the same procedure as in deriving Eq.~(8). Finally, at time $t_{\rm B}$ the hard-wall-cell potential is switched off and $U({\mathbf x}; \Gamma_{\rm B}, L_{\rm B})$ is switched on. This gives that
\begin{equation}
e^{-\beta \Delta F}=r^N\langle e^{\beta [U({\mathbf x}; \Gamma_{\rm A}, L_{\rm A})-U(r{\mathbf x}; \Gamma_{\rm B}, L_{\rm B})]}\rangle_{{\mathrm A},{\mathbf x}},
\end{equation}
where $r=L_{\rm B}/L_{\rm A}$. For $L_{\rm B}=L_{\rm A}$ it reduces to Eq.~(12).

\section{Extension to 2D and 3D cases}

Our general scheme based on Eq.~(7) does not depend on the system dimension, which can be seen from its establishment in Sec.~II. Therefore it can be applied to 2D and 3D systems as well. Nevertheless, as 2D and 3D systems are more complicated, in general it would be more challenging to design an appropriate virtual integrable system to simplify the calculation of FED. Taking the volume change problem as an example, for a 2D or 3D lattice system, its shape can also change as the volume if a twist force is exerted. In this case, the hard-wall-cell potential cannot be used by simply adopting its 2D and 3D version. Hence how to design appropriate virtual integrable systems needs more study in attempting to put Eq.~(7) into more complicated applications.

On the other hand, it is worth noting that our scheme is developed based on the JE. As such its applicability is not expected to go beyond that of the JE. For example, a phase transition can happen in a 2D and 3D system, which may cause an abrupt change in the system's structure. Whether or to what extent the JE or its necessarily generalized version can be used to capture the corresponding free energy change is still an open issue, which is also the case for our scheme. This could be interesting for future investigations.

Coming back to the volume change problem, for the simpler case that a 2D (3D) system has a rectangle (rectangular solid) shape and changes its volume under forces or pressures perpendicularly applied on each side, the corresponding free energy change can be calculated with the help of the 2D (3D) hard-wall-cell potential, given that no phase transition occurs during this process (see Appendix B for a detailed derivation). Consider the 3D case first; Suppose that at the beginning the length, width, and height of the system are, respectively, $L_{{\rm A}, x}$, $L_{{\rm A}, y}$, and $L_{{\rm A}, z}$, and the volume of the system is $V_{\rm A}=L_{{\rm A}, x} L_{{\rm A}, y} L_{{\rm A}, z}$;
at the end they become $L_{{\rm B}, x}$, $L_{{\rm B}, y}$, $L_{{\rm B}, z}$, and $V_{\rm B}$, respectively, then the FED is
\begin{equation}
e^{-\beta \Delta F}=({V_{\rm B}}/{V_{\rm A}})^N \langle e^{\beta ({U_{\rm A}-U_{\rm B}})}
\rangle_{{\mathrm A},{\mathbf r}},
\end{equation}
where $U_{\rm A}\equiv U({\mathbf x}, {\mathbf y}, {\mathbf z}; L_{{\rm A},x}, L_{{\rm A},y}, L_{{\rm A},z})$ and $U_{\rm B} \equiv U(r_x {\mathbf x}, r_y {\mathbf y}, r_z {\mathbf z}; L_{{\rm B},x}, L_{{\rm B},y}, L_{{\rm B},z})$ with $r_\alpha \equiv L_{{\rm B}, \alpha}/L_{{\rm A}, \alpha}$ ($\alpha=x, y, z$), and the distribution function for averaging is $P_{{\rm A},{\mathbf r}} \equiv e^{-\beta U_{\rm A}}/Z_{{\rm A},{\mathbf r}}$ with $Z_{{\rm A},{\mathbf r}}\equiv \int e^{-\beta U_{\rm A}} d{\mathbf r}$. Here ${\mathbf r}\equiv ({\mathbf x}, {\mathbf y}, {\mathbf z})$ is the coordinates of all particles at the beginning ($t=t_{\rm A}$). For the 2D case, Eq.~(14) also applies and keeps its form unchanged; the only change that needs be made is to drop the terms related to the $z$ coordinate in the expressions of $U_{\rm A}$, $U_{\rm B}$, and $P_{{\rm A},{\mathbf r}}$. Similarly, Eq.~(14) also incorporates the 1D case, which reduces to Eq.~(8) when the $y$ coordinate is dropped further.

\begin{figure}[!t]
\vskip-0.2cm
\includegraphics[width=8.8cm]{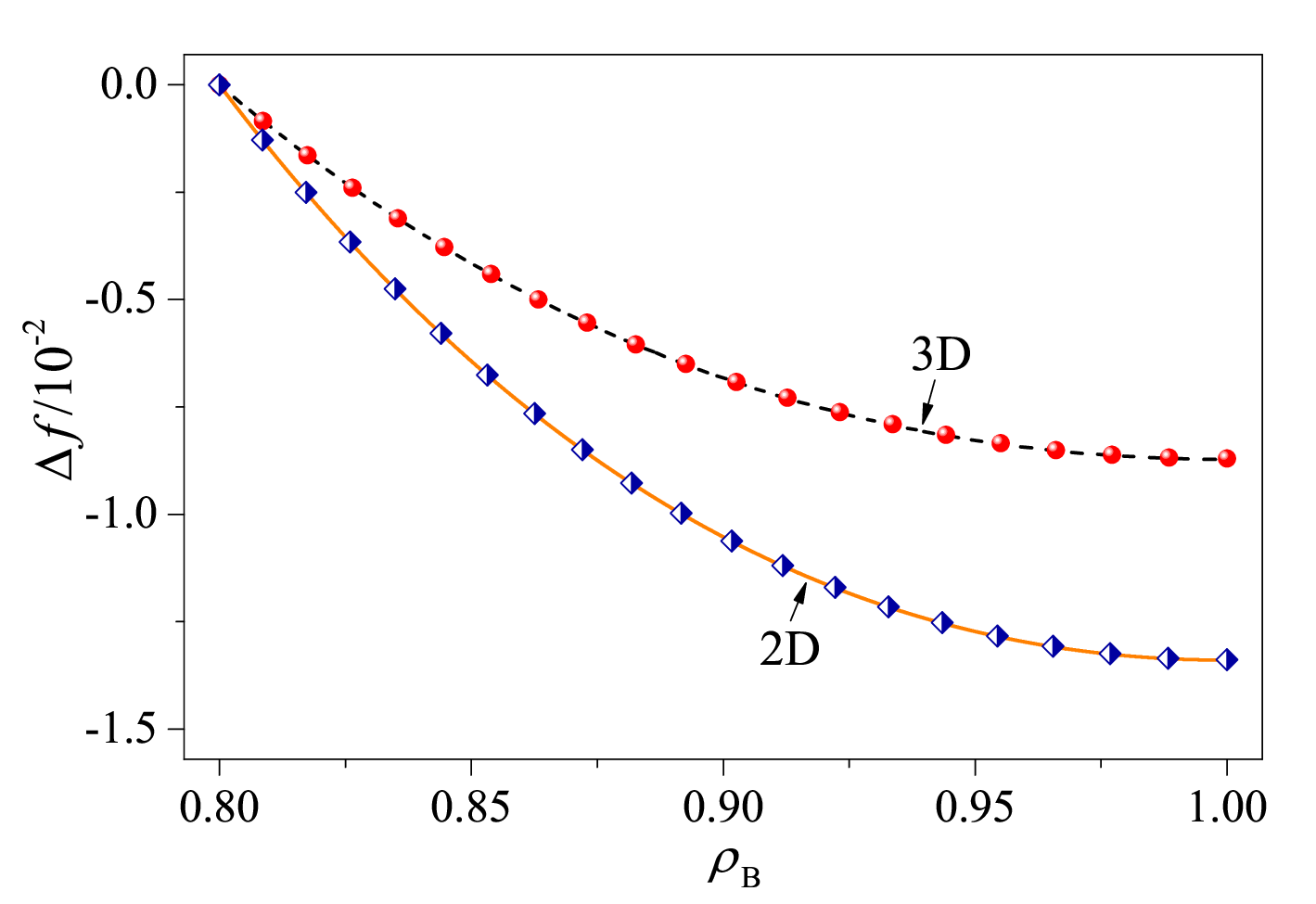}
\vskip-0.2cm
\caption{The free energy difference per particle of the 2D square (dots) and 3D cubic (diamonds) Toda lattice model with $8\times 8$ and $8\times 8\times 8$ sites, respectively, computed with our scheme Eq.~(14). The average ensemble size is 10, $\rho_{\rm A, B}\equiv N/V_{\rm A, B}$ and $\rho_{\rm A}=0.8$. The dashed and the solid line are the corresponding results of the thermodynamic integration method by integrating the numerically computed pressure (not shown).
$\beta=10^4$.}
\end{figure}

Now let us put Eq.~(14) into a numerical check. To this end, we take the square (2D) and the cubic (3D) Toda lattice with $N_x \times N_y$ and $N_x \times N_y \times N_z$ sites, respectively, as illustrating examples. The potential is
\begin{equation}
U=\sum [e^{-(|{\mathbf r}_{i}-{\mathbf r}_{j}|-1)}+(|{\mathbf r}_{i}-{\mathbf r}_{j}|-1)],
\end{equation}
where the sum runs over both $i$ and $j$ satisfying that the $i$th and the $j$th particles are the nearest neighbors and meanwhile $i<j$. The numerical results of the FED for the 2D square lattice of a square shape and that for the 3D cubic lattice of a cubic shape are shown in Fig.~7. It can be seen that again, the agreement with the benchmark is perfect.

\section{Discussions and summary}

In summary, we have explored the idea to investigate the free energy by taking advantage of a virtual system. The tremendous flexibility and possibility it implies can be envisaged, as both the Hamiltonian and the protocol can be assigned arbitrarily to some extent. Particularly, we have discussed one ``realization'' of this idea, i.e., a scheme that consists of an integrable virtual system activated (removed) simultaneously when the protocol begins (stops). Its effectiveness and efficiency have been corroborated with numerical studies.

We emphasize that our scheme based on hard-wall-cell potential represents only one possibility. Other options of the virtual system and the protocol are worth investigating, which may lead to different results that resemble Eqs.~(8) and (14). Theoretically, we believe these results may deepen our understanding of the free energy; Numerically, they may provide more optional tools for computing the free energy. In this regard, as Eqs.~(8) and (14) have shown, its advantage (compared with the JE) is that the conventional Monte Carlo algorithm is sufficient and can be adopted directly. In fact, as the computation has reduced to a sampling problem, various techniques developed for enhancing the sampling~\cite{Chipot, Frenkel85, Frenkel} can be employed to increase its efficiency further. This could be another interesting issue to explore for future studies.

\section*{Acknowledgements}
This work is supported by NSFC under Grants No. 11535011 and No. 11335006.

\begin{appendix}

\section{Motion of a particle in a 1D cell with a moving boundary}

See Fig.~8. Consider a point particle of mass $m$ confined to move freely in a one-dimensional cell with two hard boundaries. When the particle collides with one boundary, it will be reflected back elastically. The left boundary is kept fixed and the right boundary moves at a fixed velocity, $u$. Initially, the size of the cell is $l_{\rm A}$, and the position and the velocity of the particle is $x$ and $v$, respectively. After a certain time, denoted as $\tau$, the size of the cell becomes $l_{\rm B}$. Apparently, $\tau=(l_{\rm B}-l_{\rm A})/u$. Given these, in the following we will discuss the position and the velocity of the particle, denoted as $x'$ and $v'$, at time $\tau$. Note that in Ref.~\cite{Lua05} this problem has been studied for confirming Jarzynski's equality with one-dimensional, noninteracting gas.

\begin{figure}[!t]
\vskip-0.3cm
\includegraphics[width=8.8cm]{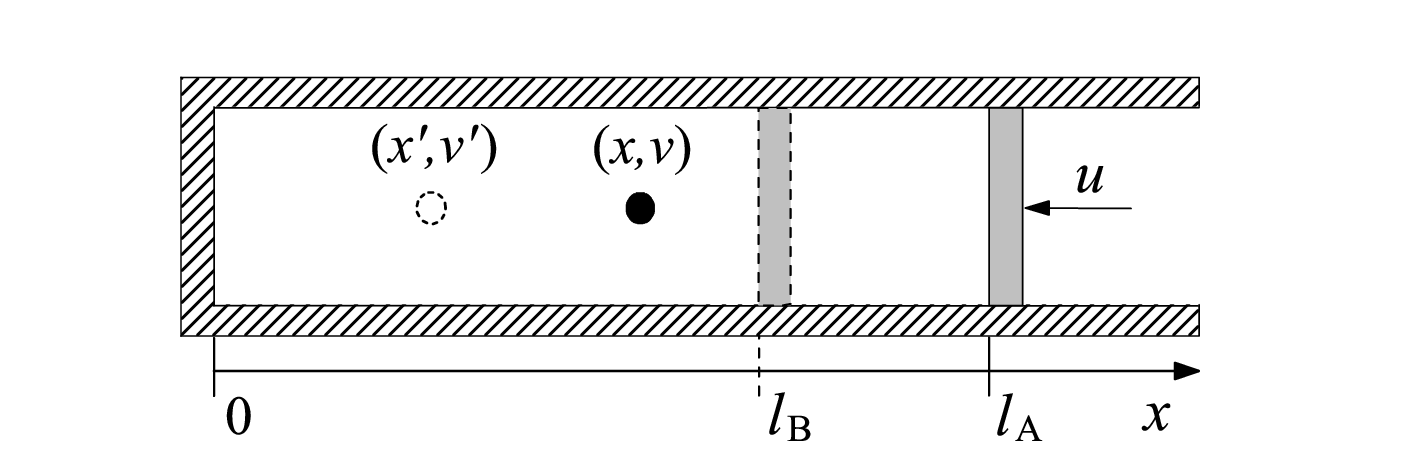}
\vskip-0.2cm
\caption{Schematic plot for the to-and-fro motion of a point particle in a cell with the right boundary moving at a fixed velocity $u$. The initial position of the right boundary is at $x=l_{\rm A}$; the initial position and velocity of the particle is $x$ and $v$. When the right boundary moves to $x=l_{\rm B}$, the position and velocity of the particle becomes $x'$ and $v'$.}
\end{figure}

Let us consider the case $u<0$, i.e., the right boundary moves to the left. The results can be extended to $u\ge 0$ straightforwardly. In this case, (a) if $0<x+v\tau<l_{\rm B}$, then the particle does not collide with any boundary during time $\tau$ and $v'=v$, $x'=x+v\tau$; Otherwise, (b) if $-l_{\rm B}<x+v\tau\le 0$, then the particle only collides with the left boundary for once, so that $v'=-v$ and $x'=-(x+v\tau)$.

Other than these two simple cases, the particle will collide with the right boundary for at least one time. (c) For $x+v\tau\ge l_{\rm B}$, right before the first collision with the right boundary, the particle's position and velocity is, respectively, $v_1=v$ and $x_1=l_{\rm A}+ut_1$, where $t_1=(l_{\rm A}-x)/(v-u)$ is the time when the first collision occurs. Similarly, (d) for $x+v\tau\le -l_{\rm B}$, we have $v_1=-v$, $x_1=l_{\rm A}+ut_1$, and $t_1=-(l_{\rm A}+x)/(v+u)$, instead.

For cases (c) and (d), it is easy to establish the map from $x_1$ and $v_1$ to the particle's state right before the $i$th collision with the right boundary that occurs at time
\begin{eqnarray}
t_i=t_1+\frac{2(i-1)x_1}{v_1-2iu+u}
\end{eqnarray}
as follows:
\begin{eqnarray}
v_i=v_1-2(i-1)u,\nonumber\\
x_i=\frac{v_1-u}{v_1-2iu+u}x_1.
\end{eqnarray}
The total number, $n$, of collisions with the right boundary during time $\tau$ satisfies $t_n<\tau<t_{n+1}$, which gives that
\begin{eqnarray}
n=1+\left[\frac{(v_1-u)(\tau-t_1)}{2l_{\rm B}}\right]_{int},
\end{eqnarray}
where the brackets represent the integer part of the variable inside. Right after the last collision, the particle's velocity becomes
\begin{eqnarray}
v_{n}^{+}=2nu-v_1.
\end{eqnarray}
Finally, for cases (c) and (d), if
\begin{eqnarray}
0<x_{n}+(\tau-t_n)v_{n}^{+},
\end{eqnarray}
then we have
\begin{eqnarray}
v'=v_n^{+},~~~~~~~~\nonumber\\
x'=x_n+(\tau-t_n)v_n^{+};
\end{eqnarray}
otherwise,
\begin{eqnarray}
v'=-v_n^{+},~~~~~~~\nonumber\\
x'=-[x_n+(\tau-t_n)v_n^{+}].
\end{eqnarray}
It follows that the total work the right boundary does to the particle during the whole process is
\begin{eqnarray}
w=\frac{1}{2}m[({v'})^{2}-v^2].
\end{eqnarray}

In the limit $u\to 0$, i.e., the right boundary moves infinitely slow, from Eqs.~(A3) and (A4) we have $nu\to v_1(l_{\rm B}-l_{\rm A})/(2l_{\rm B})$ and $v^{+}_n\to {v_1}{l_{\rm A}}/{l_{\rm B}}$, suggesting that the kinetic energy of the particle becomes $(l_{\rm A}/l_{\rm B})^2$ times that of its initial value. Therefore, the total work performed on the particle is
\begin{eqnarray}
w=\frac{1}{2}mv^2\left[\frac{l_{\rm A}^2}{l_{\rm B}^2}-1\right].
\end{eqnarray}

\section{Derivation of Eqs.~(8) and (14)}

\begin{figure*}[!t]
\vskip-0.2cm
\includegraphics[width=18cm]{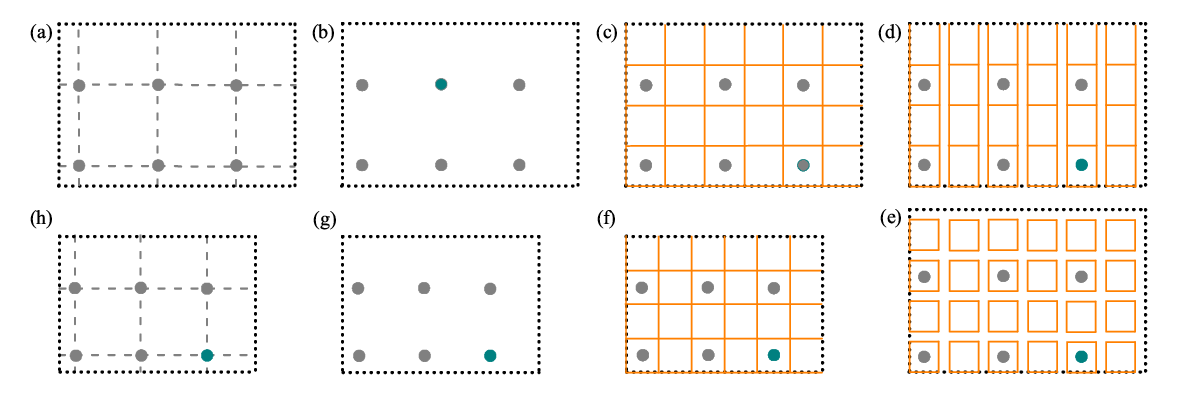}
\vskip-0.2cm
\caption{The scheme for evaluating the free energy difference of a 2D system between a reference system volume, $V_{\rm A}=L_{{\rm A},x}L_{{\rm A},y}$ (a), and a given system volume, $V_{\rm B}=L_{{\rm B},x}L_{{\rm B},y}$ (h), which is illustrated with a 2D square lattice system of $N_x\times N_y=3\times 2$ sites. The four black dotted lines in each panel represent the boundaries of the system (here the periodic boundary conditions are adopted for example). (a) For $t\le t_{\rm A}$, the interaction $U$ of the system, represented by the gray dashed lines, operates; (b) at $t=t_{\rm A}$, interaction $U({\mathbf x}(t_{\rm A}), {\mathbf y}(t_{\rm A}); L_{{\rm A},x}, L_{{\rm A},y})$ is removed and (c) a 2D hard-wall-cell potential, $V({\mathbf x}(t_{\rm A}), {\mathbf y}(t_{\rm A}); L_{{\rm A},x}, L_{{\rm A},y})$, represented by small cells, is switched on simultaneously. At this moment work $\tilde w_{\rm A}$ is calculated. For $t_{\rm A}<t<t_{\rm B}$, each cell is shrunk (d) by moving its right boundary at velocity $u_x$ first, then (e) by moving its top boundary at velocity $u_y$. During this process work $\tilde w_{V}$ is evaluated. (f) At $t=t_{\rm B}$, cells are aligned one by one first, then (g) interaction $V({\mathbf x}(t_{\rm B}), {\mathbf y}(t_{\rm B}); L_{{\rm B},x}, L_{{\rm B},y})$ is removed and (g) the original interaction $U({\mathbf x}(t_{\rm B}), {\mathbf y}(t_{\rm B}); L_{{\rm B},x}, L_{{\rm B},y})$ is switched back. At this moment work $\tilde w_{\rm B}$ is evaluated.}
\end{figure*}

Here we calculate $\tilde w_{V}$ in the virtual system with the hard-wall-cell potential and substituting the result into Eq.~(7), i.e.,
\begin{equation}
e^{-\beta \Delta F}=\langle e^{-\beta (\tilde w_{\rm A}+\tilde w_{\rm B}+\tilde w_{V})}\rangle_{\mathrm A},
\end{equation}
to obtain Eq.~(8) for the 1D case and Eq.~(14) for the 2D and 3D cases.

For the 1D case, when the system changes its volume from $L_{\rm A}$ to $L_{\rm B}$ (see Fig.~1), the hard-wall cell a particle resides in changes its volume from $l_{\rm A}=L_{\rm A}/N_c$ to $l_{\rm B}=L_{\rm A}/N_c$. According to Eq.~(A9), in the limit that $u\to 0$, the work done to a particle, say the $i$th, during this process is $\tilde w_{V, i}=\frac{1}{2}m_i v_i^2(t_{\rm A}) (\frac{1} {r^2} -1)$ with $r=L_{\rm B}/L_{\rm A}=l_{\rm B}/l_{\rm A}$. As a result,
\begin{equation}
\tilde w_V = \sum \tilde w_{V,i} =(\frac{1}{r^2}-1)\sum \frac{p_i^2(t_{\rm A})}{2m_i} .
\end{equation}
On the other hand, if we take further the limit that $N_c \to \infty$, then the $i$th particle changes its position from $x_i(t_{\rm A})$ to $x_i(t_{\rm B})=r x_i(t_{\rm A})$, implying that for the whole system, the coordinates change from ${\mathbf r}(t_{\rm A})$ to ${\mathbf r}(t_{\rm B}) = {\mathbf x}(t_{\rm B}) =r {\mathbf x}(t_{\rm A})=r {\mathbf r}(t_{\rm A})$, which leads to [see Eq.~(6)] $\tilde w_{\rm A}=-U({\mathbf x}(t_{\rm A}); \lambda_{\rm A})$ and $\tilde w_{\rm B}=U({\mathbf x}(t_{\rm B}); \lambda_{\rm B})=U(r{\mathbf x}(t_{\rm A}); \lambda_{\rm B})$,  considering that for the hard-wall-cell potential we have $V({\mathbf x}(t_{\rm A}); \lambda_{\rm A})=V({\mathbf x}(t_{\rm B}); \lambda_{\rm B})=0$. Now, by substituting $\tilde w_{\rm A}$, $\tilde w_{\rm B}$, and $\tilde w_V$ into Eq.~(7), we have
\begin{equation}
e^{-\beta \Delta F}=\frac {\int{e^{-\frac{\beta}{r^2}\sum \frac{p_i^2}{2m_i}}d{\mathbf p}}}{\int{e^{-\beta\sum \frac{p_i^2}{2m_i}}d{\mathbf p}}}\cdot\frac{\int{e^{-\beta U(r{\mathbf x};\lambda_{\rm B})}d{\mathbf x}}}{\int{e^{-\beta U({\mathbf x};\lambda_{\rm A})}d{\mathbf x}}},
\end{equation}
where the product of the two denominators on the r.h.s. is the partition function of state A (with system volume $L_{\rm A}$). The first term on the r.h.s. can be integrated out, which equals $r^N$, and the second term can be expressed as the ensemble average over distribution function $P_{{\rm A},{\mathbf x}} \equiv e^{-\beta U({\mathbf x};\lambda_{\rm A})}/Z_{{\rm A},{\mathbf x}}$ with $Z_{{\rm A},{\mathbf x}}\equiv \int e^{-\beta U({\mathbf x};\lambda_{\rm A})} d{\mathbf x}$. Then we have
\begin{equation}
e^{-\beta \Delta F}=r^N\langle e^{\beta [U({\mathbf x}; L_{\rm A})-U(r{\mathbf x}; L_{\rm B})]}\rangle_{{\mathrm A},{\mathbf x}},
\end{equation}
which is exactly Eq.~(8).

Next, let us deal with the 2D case. A schematic illustration of our scheme is presented in Fig.~9. Following the same line as in the 1D case, when the system changes its volume from $V_{\rm A}=L_{{\rm A},x}L_{{\rm A},y}$ to $V_{\rm B}=L_{{\rm B},x}L_{{\rm B},y}$, the length and the width of the hard-wall cells change from $l_{{\rm A},x}=L_{{\rm A},x}/N_{c,x}$ and $l_{{\rm A},y}=L_{{\rm A},y}/N_{c,y}$ to $l_{{\rm B},x}=L_{{\rm B},x}/N_{c,x}$ and $l_{{\rm B},y}=L_{{\rm B},y}/N_{c,y}$, respectively. Here $N_{c,x}$ and $N_{c,y}$ are the number of cells in $x$ and $y$ direction, respectively. This process can be divided into two steps: First, the cells are pressed in the $x$ direction by moving their right boundaries at a speed $u_x$ [see Fig.~9(d)]. Based on Eq.~(A9), at the limit that $u_x\to 0$, the work done to the $i$th particle is $\frac{1}{2}m_i v_{i,x}^2(t_{\rm A}) (\frac{1} {r_x^2} -1)$, where $r_x=L_{{\rm B},x}/L_{{\rm A},x}=l_{{\rm B},x}/l_{{\rm A},x}$. Note that as the motion of a particle in a rectangular cell is independent in the $x$ and $y$ directions, this result is independent of the particle's state component in the $y$ direction. Next, the cells are pressed in the $y$ direction by moving their top boundaries at a speed $u_y$ [see Fig.~9(e)]. Again, based on Eq.~(A9), at the limit that $u_y\to 0$, the work done to the $i$th particle reads $\frac{1}{2}m_i v_{i,y}^2(t_{\rm A}) (\frac{1} {r_y^2} -1)$ with $r_y=L_{{\rm B},y}/L_{{\rm A},y}=l_{{\rm B},y}/l_{{\rm A},y}$. Similarly, this part of the work has nothing to do with the particle's state component in the $x$ direction. To sum all the work done to all the particles, we have that
\begin{equation}
\tilde w_V = (\frac{1}{r_x^2}-1)\sum \frac{p_{i,x}^2(t_{\rm A})}{2m_i}+(\frac{1}{r_y^2}-1)\sum \frac{p_{i,y}^2(t_{\rm A})}{2m_i} .
\end{equation}

As to $\tilde w_{\rm A}$ and $\tilde w_{\rm B}$, as in the limits $N_{c,x}\to \infty$ and $N_{c,y}\to \infty$ we have $x_i(t_{\rm B})=r_x x_i(t_{\rm A})$ and $y_i(t_{\rm B})=r_y y_i(t_{\rm A})$, i.e., ${\mathbf x}(t_{\rm B}) =r_x {\mathbf x}(t_{\rm A})$ and ${\mathbf y}(t_{\rm B}) =r_y {\mathbf y}(t_{\rm A})$; we can write them down immediately [see Eq.~(6)]: $\tilde w_{\rm A}=-U_{\rm A}$ and $\tilde w_{\rm B}=U_{\rm B}$, where $U_{\rm A}= U({\mathbf x}, {\mathbf y}; L_{{\rm A},x}, L_{{\rm A},y})$ and $U_{\rm B} = U(r_x {\mathbf x},r_y {\mathbf y}; L_{{\rm B},x}, L_{{\rm B},y})$. Finally, by substituting $\tilde w_{\rm A}$, $\tilde w_{\rm B}$, and $\tilde w_V$ into Eq.~(7), we obtain Eq.~(14), i.e,
\begin{equation}
e^{-\beta \Delta F}=({V_{\rm B}}/{V_{\rm A}})^N \langle e^{\beta ({U_{\rm A}-U_{\rm B}})}
\rangle_{{\mathrm A},{\mathbf r}}
\end{equation}
with the distribution function for averaging being $P_{{\rm A},{\mathbf r}} = e^{-\beta U_{\rm A}}/Z_{{\rm A},{\mathbf r}}$. Note that this result can be extended to the 3D case straightforwardly.

\end{appendix}

\end{document}